\documentclass[preprintnumbers, 
endnote,nofootinbib,prl,11pt,superscriptaddress]{article}

\usepackage{epsfig,subfigure}

\usepackage{epstopdf}
\usepackage[table]{xcolor}

\usepackage[utf8]{inputenc}
\usepackage{graphicx}
\usepackage{amssymb}
\usepackage{amsxtra}
\usepackage{amsmath}
\usepackage{booktabs,multirow,tabularx}
\usepackage{slashed}
\usepackage{float}
\usepackage{placeins}
\usepackage{rotating}
\usepackage{lscape}
\usepackage{color}
\usepackage{hyperref}
\usepackage{enumitem}
\usepackage{ulem}

\usepackage[Q=yes,pverb-linebreak=no]{examplep}

\makeatletter
\newcommand{\specialcell}[1]{\ifmeasuring@#1\else\omit$\displaystyle#1$\ignorespaces\fi}
\makeatother

\DeclareGraphicsExtensions{.eps}
\graphicspath{{./}}
\newcommand{\pt}{p_{\scriptscriptstyle T}}

\newcommand{\be}{\begin{equation}}
\newcommand{\ee}{\end{equation}}

\newcommand{\mt}{m_t}

\newcommand{\nn}{\nonumber}

\newcommand\sss{\scriptscriptstyle}

\newcommand{\gev}{\,\textrm{GeV}}

\newcommand{\mh}{m_{ \sss H}}
\newcommand{\mz}{m_{ \sss Z}}
\newcommand{\dm}{\Delta_m}
\newcommand{\Gfer}{G_{ \sss F}}

\def\beq{\begin{equation}}
\def\bea{\begin{eqnarray}}
\def\eeq{\end{equation}}
\def\eea{\end{eqnarray}}
\def\beqnl{\begin{align}}
\def\endal{\end{align}}

\newcommand\amp{{\cal A}}

\newcommand\as{\alpha_{\sss S}}

\newenvironment{appendletterA}
 {
  \typeout{ Starting Appendix \thesection }
 \renewcommand{\thesection}{\Alph{section}}
  \setcounter{section}{0}
  \setcounter{equation}{0}
  
 }{
  \typeout{Appendix done}
 }

 \newenvironment{appendletterB}
 {
  \typeout{ Starting Appendix \thesection }
 \renewcommand{\thesection}{\Alph{section}}
  \setcounter{section}{1}
  \setcounter{equation}{0}
  
 }{
  \typeout{Appendix done}
 }

\makeatletter
\newcommand{\normalorbold}{%
  \ifnum\pdf@strcmp{\math@version}{bold}=\z@ bx\else m\fi
}
\makeatother

\begin{document}\color{black}
\begin{titlepage}
\nopagebreak
{\flushright{
        \begin{minipage}{5cm}
      HU-EP-21/05-RTG
        \end{minipage}        }

}
\renewcommand{\thefootnote}{\fnsymbol{footnote}}
\vspace{1cm}
\begin{center}
  {\Large \bf \color{magenta} Virtual corrections to $gg\to ZH$ via a
    transverse momentum  expansion}
  
\bigskip\color{black}\vspace{0.6cm}
     {\large\bf Lina Alasfar$^a$\footnote{email: alasfarl@physik.hu-berlin.de},
       Giuseppe Degrassi$^b$\footnote{email: giuseppe.degrassi@uniroma3.it},
       Pier Paolo Giardino$^{c}$\footnote{email: pierpaolo.giardino@usc.es},
       Ramona Gr\"{o}ber$^{d}$\footnote{email: ramona.groeber@pd.infn.it},
       Marco Vitti$^b$\footnote{email: marco.vitti@uniroma3.it} }
     \\[7mm]
{\it (a) Humboldt-Universit\"at zu Berlin, Institut f\"ur Physik, D-12489 Berlin, Germany}\\[1mm]     
{\it (b) Dipartimento di Matematica e Fisica, Universit{\`a} di Roma Tre and \\
 INFN, sezione di Roma Tre, I-00146 Rome, Italy}\\[1mm]
{\it (c) Instituto Galego de F\'isica de Altas Enerx\'ias,
  Universidade de Santiago de Compostela, 15782 Santiago de Compostela,
  Galicia-Spain}\\[1mm]
{\it (d) Dipartimento di Fisica e Astronomia 'G.~Galilei', Universit\`a di Padova and INFN, sezione di Padova, I-35131 Padova, Italy}\\     
\end{center}

\bigskip
\bigskip
\bigskip
\vspace{0.cm}

\begin{abstract}
  We compute the next-to-leading virtual QCD corrections to the partonic cross
  section of the production of a Higgs boson in association with a $Z$ boson
  in gluon fusion. The calculation is based on the recently introduced method
  of evaluating  the  amplitude via an expansion in terms of a small
  transverse momentum. We generalize the method to the case of different
  masses in the final state and of  a process not symmetric in the
  forward-backward direction exchange.  Our analytic approach gives a very
  good approximation (better than percent) of the partonic
  cross section
  in the center of mass energy region up to $\sim 750 \gev$, where
  at the LHC
  $\sim 98\%$ of the
   total hadronic cross section is concentrated.

\end{abstract}
\vfill  
\end{titlepage}    

\setcounter{footnote}{0}

\section{Introduction}
The study of the characteristics of the Higgs boson is one of the
primary tasks of the LHC program: the forthcoming Run3 and the High-Luminosity
phase will increase the
accuracy in the measurement of Higgs production cross sections and
decay rates, allowing for a more stringent test of the Standard Model
(SM) predictions.  One of the main production processes being
investigated is the so called Higgs-strahlung process $pp\to V H$, in
which a single Higgs boson is emitted together with a weak vector
boson ($V=Z, W$). The leptonic decays of the weak boson can be
exploited as a trigger for measurements of elusive Higgs decays. In
particular, the decay $H \rightarrow b \bar{b}$ has been 
observed for the first time by ATLAS and CMS, using an analysis
focused precisely on the associated production category
\cite{Aaboud:2018zhk, Sirunyan:2018kst}.

In this paper, we are interested in the associated production of a
Higgs and a $Z$ boson.  The theoretical predictions for $p p \rightarrow
Z H$ are accurate at next-to-next-to-leading-order (NNLO) in QCD, and
at next-to-leading-order (NLO) in the EW interactions \cite{Amoroso:2020lgh}.
The leading and next-to-leading contributions are connected to the
$q \bar{q}$-initiated channel, allowing to interpret $p p \rightarrow Z H$
mainly as a Drell-Yan process \cite{Han:1991ia,Brein:2003wg}.

The gluon-initiated channel $g g \rightarrow Z H$ arises for the first time
at NNLO in QCD. It is an $\mathcal{O}(\as^2)$ correction, but the
contribution from this process to the hadronic cross section is
non-negligible because of the large gluon luminosity at the LHC.  It
has been shown that the relevance of $g g \rightarrow Z H$ is even
more enhanced in the boosted kinematic regime, to the point of being
comparable to the quark-initiated contribution near the $t \bar{t}$
threshold \cite{Englert:2013vua}.  The factorization- and
renormalization-scale uncertainties related to the gluon-induced
process also affect significantly the uncertainty on the total $p p
\rightarrow Z H$ cross section. This issue is specific to the $Z
H$ final state, since the gluon-induced channel is absent in $p p
\rightarrow W H$. The knowledge of the NLO corrections to $g g
\rightarrow Z H$ would reduce the scale uncertainties, facilitating
precision studies in the next runs of the LHC. The $g g \rightarrow Z H$
contribution is relevant also for New
Physics (NP) studies, since it is sensible to both sign and magnitude
of the top Yukawa coupling,  dipole operators \cite{Englert:2016hvy}
 and can receive additional contributions
from new particles \cite{Harlander:2013mla}.

An improved knowledge of the SM prediction for the gluon-induced
contribution is therefore very important both for precision
measurements of $Z H$ production within the SM and for testing NP in
this channel.  The leading order (LO) contribution to the $g g \rightarrow Z H$
amplitude, given by one-loop diagrams, was computed exactly
in refs.\cite{Kniehl:1990iva, Dicus:1988yh}. At the NLO 
the virtual correction part contains two-loop multi-scale integrals that
constitute, at present, an obstacle to an exact evaluation of the NLO
contribution. Specifically, the corrections due to the
two-loop box diagrams are still not known analytically.  A first
computation of the NLO terms was obtained in
ref.\cite{Altenkamp:2012sx} using an asymptotic expansion in the limit
$\mt \rightarrow \infty$ and $m_b = 0$, and pointed to a $K$-factor of
about 100\% with respect to the LO contribution. Soft gluon resummation has
been performed in ref.\cite{Harlander:2014wda} including next-to-leading
logarithmic terms, and the result has been matched to the fixed NLO
computation of ref.\cite{Altenkamp:2012sx}.  Finite top-quark-mass effects
to $g g \rightarrow Z H$ have been investigated in
ref.\cite{Hasselhuhn:2016rqt} using a combination of large-$\mt$ expansion
(LME) and Pad\'e approximants.  In addition, a data-driven method to extract
the non-Drell-Yan part of $p p \rightarrow Z H$, which is dominated by
the gluon-induced contribution, has been proposed in
ref.\cite{Harlander:2018yns}, exploiting the known relation between $W H$
and $ Z H$ associated production when only the Drell-Yan component of
the two processes is considered.  A qualitative study focusing on
patterns in the differential distribution has been conducted in
ref.\cite{Hespel:2015zea}, where $2 \rightarrow 2$ and $2 \rightarrow 3$
LO matrix elements were merged and matched to improve the description
of the kinematics.

Very recently, a new analytic computation  of the NLO virtual contribution
based on a high-energy expansion of the amplitude, supported by Pad\'e
approximants, and on an improved LME, has been
carried out \cite{Davies:2020drs}. The results are in agreement with a new
exact numerical study \cite{Chen:2020gae}, in the energy regions where the
expansions are legitimate. Nonetheless, an improvement on the analytic
calculation is still desirable, since the heavy-top and the
high-energy expansions do not cover well the region
$350\, {\rm GeV}\lesssim \sqrt{\hat{s}} \lesssim 750\, {\rm GeV}$, where
$\sqrt{\hat{s}}$ is the partonic center of mass energy. It should be
remarked  that this region provides a significant part of the hadronic cross
section at the LHC, about 68\%.

In this paper, we present an analytic calculation of the virtual NLO QCD
corrections to the $gg\to Z H$ process that covers the region
$\sqrt{\hat{s}} \lesssim 750\, {\rm GeV}$,  which contributes about 98\% to the hadronic
cross section. The most difficult parts, i.e.~the two-loop box
diagrams, are computed in
terms of a forward kinematics  \cite{Bonciani:2018omm} via an  expansion in
the $Z$ (or Higgs) 
transverse momentum, $\pt$, while the rest of the
virtual corrections is computed exactly. 
We remark that our calculation is complementary
to the results of ref.\cite{Davies:2020drs}, which covers the region of
large transverse momentum of the $Z$. Furthermore, the merging of the two
analyses
allows an analytic evaluation of the NLO virtual corrections in
$gg\to Z H$ in the entire phase space.

The paper is structured as follows: in the next section we introduce our
notation and the definitions of the form factors in terms of which we
express the amplitude. In \hyperref[sec:tre]{section \ref*{sec:tre}}, we present  the expansion of
the
amplitude in terms of the $Z$ transverse momentum. \hyperref[sec:Add]{Section \ref*{sec:Add}}
is devoted to a discussion of the expected range of validity of
the evaluation  of the amplitude via a $\pt$-expansion, by comparing the exact
result for the LO cross section with the $\pt$-approximated one. In~\hyperref[sec:quattro]{section \ref*{sec:quattro}} we present an outline of our NLO computation, while
the~\hyperref[sec:sei]{next section} contains our NLO results. Finally we present our \hyperref[sec:conclusion]{conclusions}.
The paper is complemented by two appendices. In \hyperref[app:uno]{appendix \ref*{app:uno}}, we
report the
explicit expressions for the orthogonal projectors we employ in the calculation.
We present also  the relation between our form factors and the ones used in
ref.\cite{Davies:2020drs}. In \hyperref[app:due]{appendix \ref*{app:due}}, we report the exact
results
for the triangle and the reducible double-triangle contributions. 

\section{Definitions}
In this section we  introduce our definitions for the
calculation of the NLO QCD corrections to the associated production of
a Higgs and a $Z$ boson from gluon fusion.

The amplitude  $g^\mu_a(p_1)g^\nu_b(p_2)\to Z^\rho(p_3) H(p_4)$ can be written as
\bea
&&\amp=i \sqrt{2}\frac{\mz \Gfer \as(\mu_R)}{\pi}\delta_{ab}\epsilon^a_\mu(p_1)
\epsilon^b_\nu(p_2)\epsilon_\rho(p_3)\hat{\amp}^{\mu\nu\rho}(p_1,p_2,p_3 ),\\
&&\hat{\amp}^{\mu\nu\rho}(p_1,p_2,p_3 )=\sum_{i=1}^{6}
\mathcal{P}_i^{\mu\nu\rho}(p_1,p_2,p_3 )
\amp_i(\hat{s},\hat{t},\hat{u},\mt,\mh,\mz),
\label{eq:amp}
\eea
where $\Gfer$ is the Fermi constant, $\as(\mu_R)$ is the strong coupling
constant defined at a  scale $\mu_R$ and
$\epsilon^a_\mu(p_1)\epsilon^b_\nu(p_2)\epsilon_\rho(p_3)$ are the
polarization vectors of the gluons and the $Z$ boson, respectively. The
tensors $\mathcal{P}_i^{\mu\nu\rho}$ are a set of orthogonal
projectors, whose explicit expressions are presented in appendix \ref{app:uno}.
The corresponding form factors
$\amp_i(\hat{s},\hat{t},\hat{u},\mt,\mh,\mz)$ are functions of the
masses of the top quark ($\mt$), Higgs ($\mh$) and $Z$ ($\mz$) bosons, and of
the partonic Mandelstam variables
\beq
\hat{s}=(p_1+p_2)^2,~~ \hat{t}=(p_1+p_3)^2,~~ \hat{u}=(p_2+p_3)^2,
\eeq
where $\hat{s}+\hat{t}+\hat{u}=\mz^2+\mh^2$ and we took all the momenta to
be incoming.

The $\amp_i$ form factors can be expanded up to NLO terms as
\beq
\amp_{i} = \amp_i^{(0)} + \frac{\as}{\pi} \amp_i^{(1)}
\label{eq:ampexp}
\eeq
and the  Born partonic cross section can be written as
\beq
\hat{\sigma}^{(0)}(\hat{s})=
\frac{\mz^2 \Gfer^2 \as(\mu_R)^2}{64 \hat{s}^2(2\pi)^3}
\int^{\hat{t}^+}_{\hat{t}^-}d\hat{t}\sum_i \left|\amp_i^{(0)}\right|^2,
\eeq
where
$\hat{t}^\pm=[-\hat{s}+\mh^2+\mz^2\pm\sqrt{(\hat{s}-\mh^2-\mz^2)^2-4\mh^2\mz^2}\,]/2$.
\label{sec:due}

\begin{figure}
\begin{center}
\includegraphics[width=12cm]{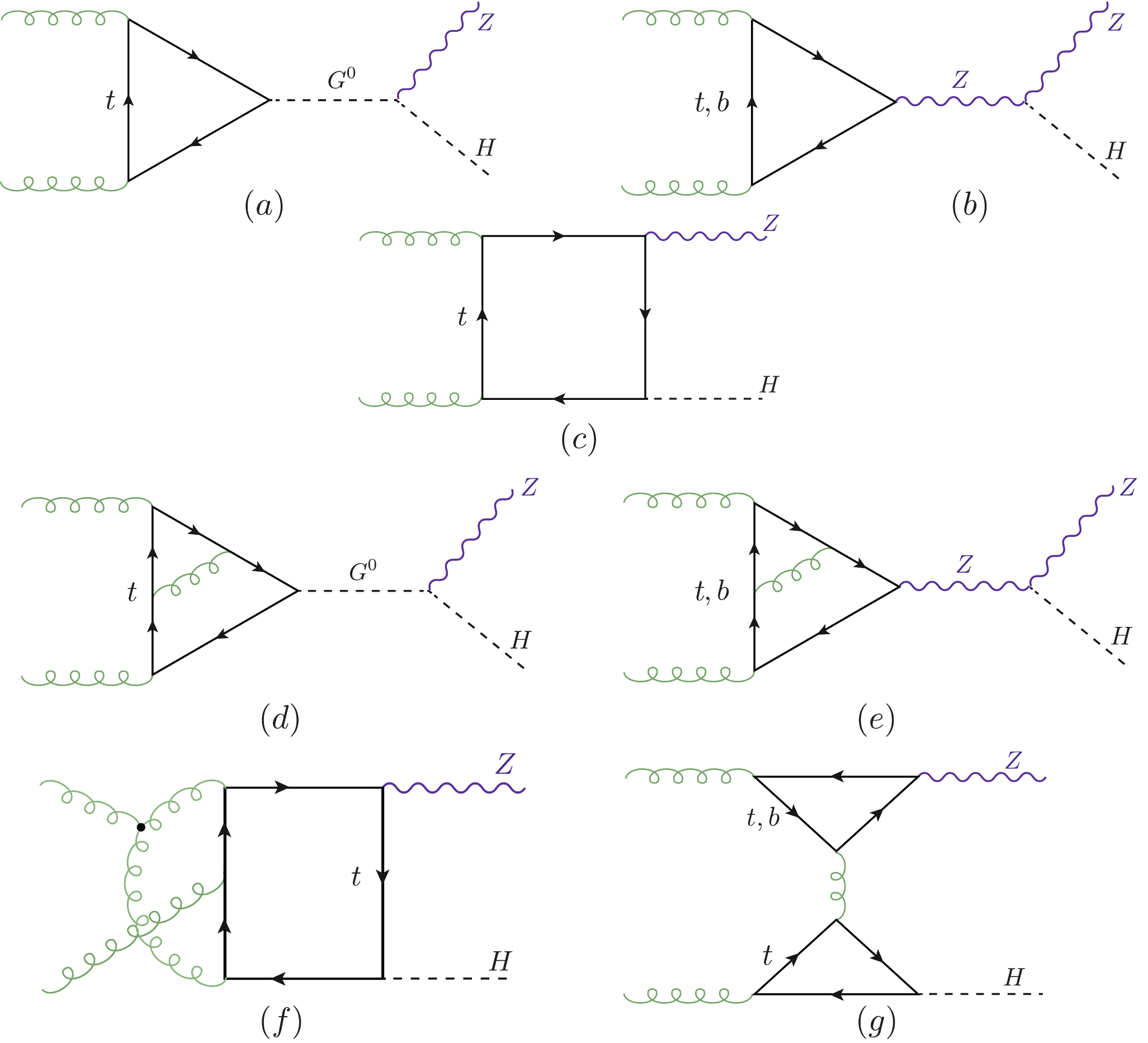}
\caption{Examples of Feynman diagrams contributing to $gg \to ZH$ at  LO and
  NLO.}
\label{fig:dia}
\end{center}
\end{figure}

The Feynman diagrams that contribute to the $gg \to  ZH$ amplitude up to NLO
can be separated into triangle, box and double-triangle
contributions, the last type appearing for the first time at the
NLO level. Examples of LO (NLO) triangle and box
categories are shown in fig.\ref{fig:dia} $(a)$ - $(c)$
($(d)$ - $(f)$).
Due to the presence of a $\gamma_5$ in the axial coupling of the $Z$ boson to
the fermions in the loop, the projectors $\mathcal{P}_i^{\mu\nu\rho}$ are
proportional to the Levi-Civita total anti-symmetric tensor
$\epsilon^{\alpha\beta\gamma\delta}$ (see appendix \ref{app:uno}),
whose treatment in dimensional regularization is, as well known, delicate
and will be discussed in section \ref{sec:quattro}.

In our calculation we treat all the quarks but the top as massless.
As a consequence, the contribution to the amplitude of the first two generations
vanishes. Concerning the third generation, the contribution of the bottom
is present  in the triangle diagrams with the exchange of a $Z$ boson
(fig.\ref{fig:dia}$(b),(e)$) and in the double-triangle diagrams
(fig.\ref{fig:dia}$(g)$).
A nice observation in ref.\cite{Altenkamp:2012sx} allows to compute
easily the full (top+bottom) triangle contribution. As noticed in that
reference,
the triangle contribution with a $Z$ exchange contains a $ggZ^*$ subamplitude
which in the Landau gauge can be related to the decay of a massive vector boson
with mass $\sqrt{\hat{s}}$ into two massless ones, a process that is
forbidden by
the Landau-Yang theorem \cite{Landau:1948kw,Yang:1950rg}. As a consequence,
the full triangle contribution can be obtained from the top triangle diagrams
with the exchange of the unphysical scalar $G^0$, with the propagator of the
$G^0$ evaluated in the Landau gauge. This part of the top triangle
  diagrams  can be obtained from the decay
amplitude of a pseudoscalar boson into two gluons which is  known in
the literature in the full mass dependence up to NLO terms \cite{Spira:1995rr,Aglietti:2006tp}. 

Given the above observation, our calculation of the NLO corrections to
the $gg \to ZH$ amplitude focuses on the analytic evaluation of the
double-triangle (fig.\ref{fig:dia}$(g)$) and two-loop box contributions
(fig.\ref{fig:dia}$(f)$). The former contribution is evaluated exactly.
The latter is evaluated via two different expansions: i) via  a LME, following
ref.\cite{Degrassi:2010eu}, up to and including ${\cal O}(1/\mt^6)$ terms,
which is expected to work below the $2\, \mt$ threshold; ii) via an expansion in
terms of the $Z$ transverse momentum, following ref.\cite{Bonciani:2018omm},
whose details are presented in the next section.

\section{Expansion in the transverse momentum}
\label{sec:tre}
The transverse momentum of the $Z$  boson can be written in
terms of the Mandelstam variables as
\beq
\pt^2=\frac{\hat{t}\hat{u}-\mz^2\mh^2}{\hat{s}}.
\label{ptdef}
\eeq
From  eq.(\ref{ptdef}), together with the relation between
the Mandelstam variables, one finds 
\beq
\pt^2+\frac{\mh^2+\mz^2}{2}\leq\frac{\hat{s}}{4}+\frac{\dm^2}{\hat{s}},
\label{ptexp}
\eeq
where
$\dm = (\mh^2 -\mz^2)/2$. Eq.(\ref{ptexp}) implies 
$\pt^2/\hat{s} < 1$ that, together with the kinematical constraints
$\mh^2/\hat{s}< 1$ and
$\mz^2/\hat{s} < 1$,  allows the expansion of the amplitude in terms of these
three ratios.

A direct expansion in $\pt$ is not possible at amplitude level, since $\pt$
itself does not appear in the amplitudes. However, as we argued in
ref.\cite{Bonciani:2018omm}, the expansion in $\pt^2/\hat{s}\ll 1$ is equivalent
to an expansion in terms of the ratio of the reduced Mandelstam variables
$t^\prime/s^\prime\ll 1$ or $u^\prime/s^\prime\ll 1$, depending whether we are
considering the process to be in a forward or backward kinematics. The
$s^\prime,\,t^\prime$ and $u^\prime$  variables are defined as
\beq
s^\prime=p_1\cdot p_2=\frac{\hat{s}}{2},~~
t^\prime=p_1\cdot p_3=\frac{\hat{t}-\mz^2}{2},~~ u^\prime =
p_2\cdot p_3=\frac{\hat{u}-\mz^2}{2}
\eeq
and satisfy
\beq
s^\prime + t^\prime + u^\prime =\dm.
\eeq

The cross section of a $2 \to 2$ process can always be expanded into
a forward and backward contribution. Looking at the dependence of $\sigma$
upon $t^\prime,\, u^\prime$ we can write
\bea
\sigma&\propto&\int^{t_f}_{t_i}dt^\prime\mathcal{F}(t^\prime,u^\prime)=
\int^{t_m}_{t_i}dt^\prime\mathcal{F}(t^\prime,u^\prime)+
\int^{t_f}_{t_m}dt^\prime\mathcal{F}(t^\prime,u^\prime) \nn \\
&\sim&\int^{t_m}_{t_i}dt^\prime\mathcal{F}(t^\prime\sim0,u^\prime\sim-s^\prime)+
\int^{t_f}_{t_m}dt^\prime \mathcal{F}(t^\prime\sim -s^\prime,u^\prime\sim0)
\label{eq:forback}
\eea
where $t_i=(\hat{t}^--\mz^2)/2$, $t_f=(\hat{t}^+-\mz^2)/2$ and $t_m$ is the
value of $t^\prime$ at which
$t^\prime =u^\prime=(-s^\prime+\dm)/2$. The two terms in the second
line of eq.(\ref{eq:forback}) represent the expansion in the forward and
backward kinematics, respectively.\\
If the amplitude is symmetric under $t^\prime\leftrightarrow u^\prime$
exchange then
\bea
\sigma&\propto&\int^{t_m}_{t_i}dt^\prime\mathcal{F}(0,-s^\prime)+
\int^{t_f}_{t_m}dt^\prime\mathcal{F}(-s^\prime,0)= \nn \\
&& 
\int^{t_m}_{t_i}dt^\prime\mathcal{F}(0,-s^\prime)+
\int^{t_f}_{t_m}dt^\prime\mathcal{F}(0,-s^\prime)=
\int^{t_f}_{t_i}dt^\prime\mathcal{F}(0,-s^\prime)
\eea
so that the expansion in the forward kinematics actually covers the entire
phase space.

In the case of $gg \to ZH$ the process itself is not symmetric
under the $t^\prime\leftrightarrow u^\prime$ exchange. However, as
can be seen from the explicit expressions of the projectors in appendix
\ref{app:uno},  it can be written as a sum of symmetric and antisymmetric
form factors. To perform only the expansion in the forward kinematics
one can proceed in the following way.
On the symmetric form factors the expansion can be directly performed. 
For the antisymmetric ones,
it is sufficient first  to extract  the overall antisymmetric factor
$(\hat{t}-\hat{u})$  just by multiplying the form factor by $1/(\hat{t}-\hat{u})$,
written as $1/( 2 s^\prime - 4 t^\prime - 2 \dm)$, 
then perform the expansion in the forward
kinematics and finally multiply back by $(\hat{t}-\hat{u})$.

As discussed in ref.\cite{Bonciani:2018omm}, to implement the $\pt$-expansion
at the level of Feynman diagrams it is convenient
to introduce the  vector $r^\mu = p_1^\mu +p_3^\mu$, which satisfies
\beq
r^2= \hat{t},~~ r\cdot p_1=\frac{\hat{t}-\mz^2}{2},~~
r\cdot p_2=-\frac{\hat{t}-\mh^2}{2},
\label{rsp}
\eeq
and therefore can be also written as
\beq
r^\mu =-\frac{\hat{t}-\mh^2}{\hat{s}}p_1^\mu +
\frac{\hat{t}-\mz^2}{\hat{s}} p_2^\mu + r_\perp^\mu =
\frac{t^\prime}{s^\prime}\,(p_2^\mu -p_1^\mu) - \frac{\dm}{s^\prime} \, p_1^\mu +
r_\perp^\mu,
\label{rpp}
\eeq
where
\beq
r_\perp^2=-\pt^2.
\eeq

From eq.(\ref{ptdef}) one obtains
\beq
t^\prime = -\frac{s^\prime}2 \left\{ 1 - \frac{\dm}{s^\prime} \pm
\sqrt{\left( 1 - \frac{\dm}{s^\prime} \right)^2 -
  2 \frac{\pt^2 + \mz^2}{s^\prime}} \right\}
\label{tpdef}
\eeq
that implies  that the expansion in
small $\pt$ (the minus sign case in eq.(\ref{tpdef})) can be realized
at the level of Feynman diagrams, by expanding the propagators
in terms of the vector $r^\mu$ around $r^\mu \sim 0$ or, equivalently,
$p_3^\mu \sim -p_1^\mu$, see eq.(\ref{rpp}). 

The outcome  of the evaluation of the $gg \to ZH$ amplitude via a
$\pt$-expansion is expressed in terms of a series of Master Integrals (MIs)
that are functions of $\hat{s}$ and $\mt^2$ only, and whose coefficients can be
organized in terms of powers of ratios of small  over large parameters
where $\pt^2, \, \mh^2$ and $\mz^2$ are identified as the small parameters while
$\mt^2$ and $\hat{s}$ as the large ones. 
Thus, the range of validity of the expansion depends
on  the condition that $\pt^2$ can be treated as a ``small parameter'' with
respect to $\mt^2$ because all the other ratios, small over
large, are always smaller than 1.

\section{LO Comparison}
\label{sec:Add}
In order to investigate  the range of validity of the evaluation of the
$gg \to ZH$ amplitude via a $\pt$-expansion,  we compare
the exact result for the LO partonic cross
section \cite{Kniehl:1990iva, Dicus:1988yh} with the result obtained
via our $\pt$-expansion. The latter is expressed in terms of the same four MIs
that enter into the analogous calculation of the
$gg \to HH$ LO amplitude \cite{Bonciani:2018omm}, or
\bea
B_0[\hat{s},\mt^2,\mt^2] \equiv  B_0^+, &
     B_0[- \hat{s},\mt^2,\mt^2]  \equiv B_0^- , &\\
   C_0[0,0,\hat{s},\mt^2,\mt^2,\mt^2]  \equiv  C_0^+  ,& ~~~
   C_0[0,0,-\hat{s},\mt^2,\mt^2,\mt^2]  \equiv C_0^- &
  \eea
where
\beq
B_0[q^2,m_1^2,m_2^2] = \frac1{i\pi^2}
\int \frac{d^n k}{\mu^{n-4}} \frac1{(k^2 -m_1^2)((k+q)^2-m_2^2)}
\label{Bzero}
\eeq
\begin{align}
  \specialcell{ C_0[q_a^2,q_b^2,(q_a+q_b)^2, m_1^2,m_2^2,m_3^2] = \hfill }\nn  \\
 \frac1{i\pi^2}  \int \frac{d^n k}{\mu^{n-4}} \frac1{[k^2 -m_1^2][(k+q_a)^2-m_2^2]
    [(k-q_b)^2 - m_3^2]} &
  \label{Czero}
\end{align}
are the Passarino-Veltman functions \cite{Passarino:1978jh},
with $n$ the dimension of spacetime and $\mu$ the 't Hooft mass.

As an illustration of our LO result we present the explicit expressions for 
one symmetric, ${\cal A}_2$, and one antisymmetric,
${\cal A}_6$, form factor including the first correction in the ratio of
small over large parameters which will be referred
to as\footnote{With a slight abuse of notation we indicate the
  counting of the orders in the expansion as
  $\mathcal{O}(\pt^{2n})$ that  actually means the inclusion of terms that
  scale   as $(x/y)^n$,   where $x=\pt^2,\, \mz^2,\,\mh^2$ and
   $y=\hat{s},\,\mt^2$, with respect to the $\hat{s}, \mt^2 \to \infty$
  contribution. The latter is indicated as $\mathcal{O}(\pt^0)$ and corresponds
  to the first non zero contribution in the expansion of the diagrams
  in terms of the vector $r^\mu$.} 
  ${\mathcal O}(\pt^2)$.
 We divide the result into triangle ($\triangle$) and
box ($\square$) contribution or
\bea
  \mathcal{A}_{2}^{(0, \triangle)} &=& -
  \frac{ \pt }{\sqrt{2} \left( \mz^2+\pt^2 \right)} (\hat{s}-\dm)\,
  \mt^2 C_0^+,
  \label{Adt}\\
 \mathcal{A}_{2}^{(0, \square)} &=&
\frac{ \pt }{\sqrt{2} \left(\mz^2+\pt^2 \right)}\, \Biggl\{ \Biggr. \nn \\
&& \Biggl( \mt^2 -\mz^2 \frac{ \hat{s}-6 \mt^2}{4 \hat{s}}-
\pt^2 \frac{ 12 \mt^4-16 \mt^2 \hat{s}+\hat{s}^2}{12 \hat{s}^2}
  \Biggr)  B_0^+ \nn \\
  &-& \Biggl( \mt^2 -\dm  \frac{\mt^2}{ \left( 4 \mt^2+\hat{s}\right)}
  + \mz^2 \frac{ 24 \mt^4 -6 \mt^2 \hat{s}-
    \hat{s}^2 }{4 \hat{s} \left(4 \mt^2+\hat{s}\right)} -
     \nn \\
   & & ~~~~~~~\pt^2 \frac{ 48 \mt^6-68 \mt^4 \hat{s}-4
       \mt^2 \hat{s}^2+\hat{s}^3 }{ 12 \hat{s}^2 \left(
       4 \mt^2 +\hat{s} \right) }  \Biggr)
   B_0^- \nn \\
   &+&\Biggl( 2 \mt^2- \dm +
   \mz^2 \frac{3 \mt^2-\hat{s}}{\hat{s} } +
   \pt^2  \frac{ 3 \mt^2 \hat{s}-2 \mt^4 }{\hat{s}^2}\Biggr)
   \mt^2 \, C_0^- \nn \\
   & +&\Biggl( \hat{s}-2 \mt^2 +
   \mz^2  \frac{\hat{s}-3 \mt^2 }{\hat{s}}+
   \pt^2 \frac{ 2 \mt^4-3 \mt^2 \hat{s}+\hat{s}^2}{\hat{s}^2 }\Biggr)
   \mt^2 \, C_0^+ \nn \\
   & +&\log \left(\frac{\mt^2}{\mu^2}\right) \frac{ \mt^2}{\left(4
   \mt^2+\hat{s}\right) } \Biggl( \dm + 2  \mz^2
   +\pt^2 \frac{2   \hat{s}-2 \mt^2}{3 \hat{s} }\Biggr)\nn  \\
   &-&\dm \frac{2 \mt^2}{\left(4 \mt^2+\hat{s}\right) } +
   \mz^2  \frac{\hat{s}-12 \mt^2}{4 \left(4 \mt^2+\hat{s}\right)}
   +\pt^2 \frac{ 8 \mt^4-2 \mt^2 \hat{s}+ \hat{s}^2 }
   {4\hat{s} ( 4\mt^2 + \hat{s})}  \Biggl. \Biggl\},\nn \\
   &&
   \label{Adb}
   \eea
   and
   \bea
   \mathcal{A}_{6}^{(0, \triangle)} &=&  0,
   \label{Ast} \\
\mathcal{A}_{6}^{(0, \square)} &= & 
\frac{\hat{t}-\hat{u}}{\hat{s}^2} \,\pt \Biggl[ \frac{\mt^2}2
  \Bigl( B_0^- - B_0^+ \Bigr) -\frac{\hat{s}}{4} \nn \\
   & -&\frac{2 \mt^2+\hat{s}}{2}\mt^2 \, C_0^- 
  +\frac{2 \mt^2-\hat{s}}{2} \mt^2 \,C_0^+  \Biggr],
\label{Asb}
\eea
where in eqs.(\ref{Adb},\ref{Asb}) the $B_0$ functions are understood as the
finite part of the integrals on the right hand side of eq.(\ref{Bzero}).

\begin{figure}[t]
	\centering
\includegraphics[width=\linewidth]{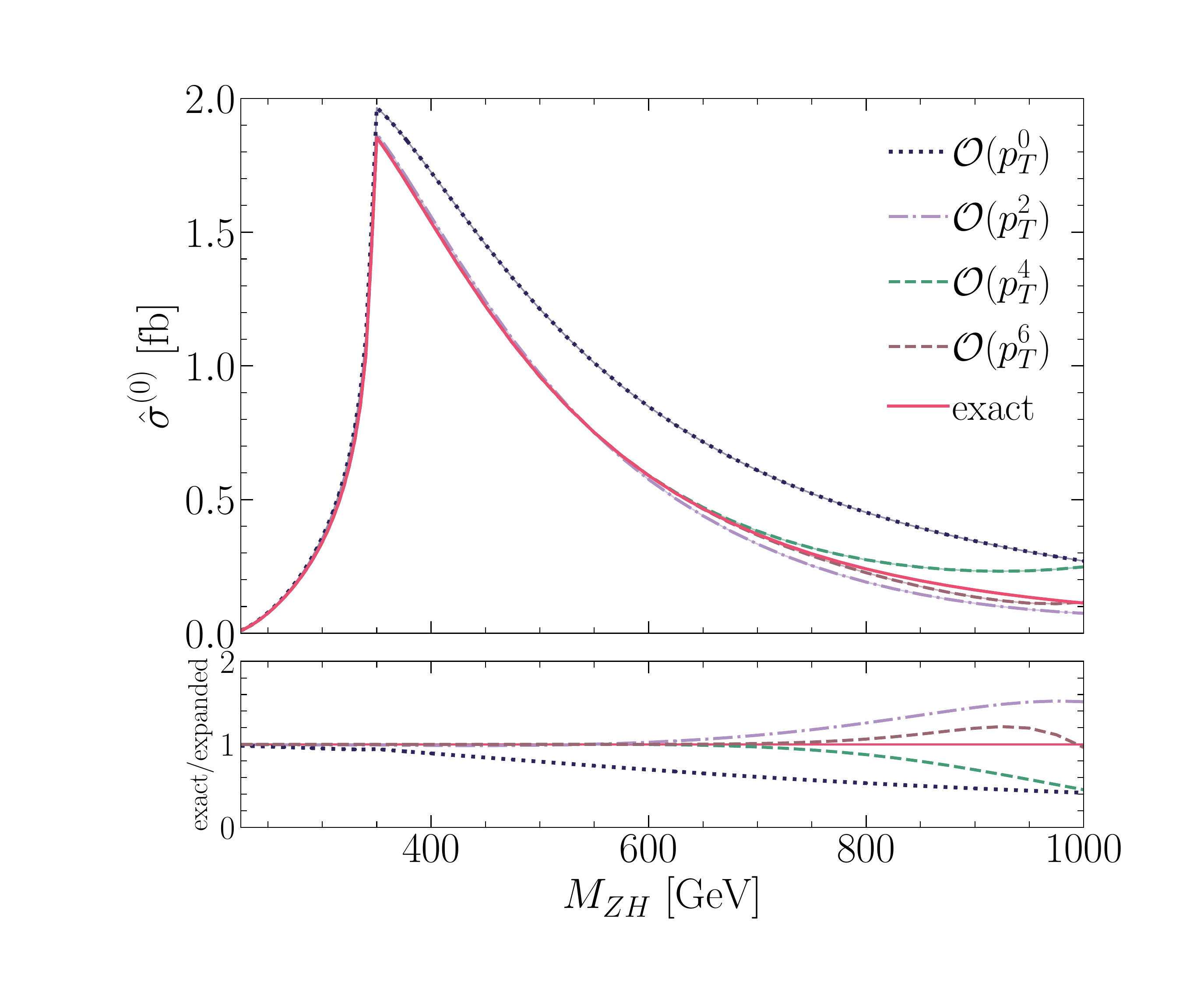}
\caption{LO partonic cross section
  as a function of the invariant mass $M_{ZH}$.
  The full result (red line) is plotted together with results at
  different orders in the $\pt$-expansion (dashed lines). In the bottom part,
  the ratio of the full result over the $\pt$-expanded one at
  various orders is shown.}
\label{fig:LO}
\end{figure}

In fig.\ref{fig:LO} the exact partonic LO cross section (red line) is
shown as a function of the invariant mass of the $ZH$ system, $M_{ZH}$, and
compared to various $\pt$-expanded results. 
For the numerical evaluation of the cross section here and in
the following, we used as SM input parameters
\begin{equation}
  \begin{split}
    \mz=91.1876 \,\gev, ~~\mh=125.1 \gev, ~~  \mt=173.21\, \gev, \nn\\
    m_b = 0 \, \gev, ~~G_F= 1.16637\,\gev^{-2},~~ \alpha_s(\mz)=0.118.
  \end{split}
\end{equation}
In the lower part of fig.\ref{fig:LO}  the ratio of the
exact result over the $\pt$-expanded one is shown.  From this ratio
one can see that 
the $\mathcal{O}(\pt^0)$ contribution covers well the $ZH$ invariant mass
region $M_{ZH}\lesssim 2\mt$, corresponding to the range of validity of an
expansion in the large top quark mass. Furthermore, when the contributions up to
$\mathcal{O}(\pt^4)$ are taken into account a remarkable agreement with the
exact result is found up to $M_{ZH}\lesssim 750\gev$.
This agreement is extended to
sligthly higher values of $M_{ZH}$ when the $\mathcal{O}(\pt^6)$ contribution is
included, a finding in close analogy to the result for di-Higgs
production \cite{Bonciani:2018omm}. Similar conclusions can be drawn from table
\ref{tab:partonic}, where it is shown that the partonic cross section
at $\mathcal{O}(\pt^4)$ agrees with the full result for
  $M_{ZH} \lesssim 600 \gev$  on the permille level 
and the agreement further improves when $\mathcal{O}(\pt^6)$ terms are included.
As a final remark for this section, we notice that, from the comparison
with the LO exact result, the $\pt$-expanded evaluation of the amplitude is  expected to provide an
accurate result up to $M_{ZH}\sim 700-750\gev$ that corresponds, from
eq.(\ref{ptexp}), to $\pt \lesssim 300-350\, \gev\approx 2\, \mt$.

\begin{table}
\renewcommand{\arraystretch}{1.2}
\centering
\begin{tabular}{| c| c | c | c| c| c|} \hline
\rowcolor{lightgray}  $M_{ZH}$ [GeV]  & $\mathcal{O}(\pt^0)$ & $\mathcal{O}(\pt^2)$ & $\mathcal{O}(\pt^4)$ & $\mathcal{O}(\pt^6)$ & full \\ \hline 
\cellcolor{lightgray} 300 & 0.3547 & 0.3393 &  0.3373 &0.3371& 0.3371 \\
\cellcolor{lightgray} 350 & 1.9385 & 1.8413& 1.8292 &1.8279& 1.8278 \\
\cellcolor{lightgray} 400 & 1.6990 & 1.5347 & 1.5161 &1.5143& 1.5142 \\
\cellcolor{lightgray} 600 & 0.8328 & 0.5653 & 0.5804 &0.5792&  0.5794 \\ 
\cellcolor{lightgray} 750 & 0.5129 & 0.2482 & 0.3129 & 0.2841 &  0.2919 \\ \hline
\end{tabular}
\caption{The partonic cross section $\hat{\sigma}^{(0)}$ at
  various orders in $\pt$ and the full computation for several values of $M_{ZH}$. \label{tab:partonic}}
\end{table}

\section{Outline of the NLO Computation}
\label{sec:quattro}
In this section we discuss our evaluation of the three different types of
diagrams that appear in the virtual corrections to the $gg \to ZH$ amplitude
at the NLO.

The triangle contribution (fig.\ref{fig:dia}$(d),(e)$) was evaluated using the
observation of  ref.\cite{Altenkamp:2012sx}, i.e.~we adapted
the result of ref.\cite{Aglietti:2006tp} for the decay of a
pseudoscalar boson into two gluons  to our case. This contribution is evaluated
exactly and explicit expressions for the form factors are presented in
appendix \ref{app:due}. We notice that if we interpret the exact result
in terms of our counting of the expansion in $\pt$, the $\pt$-expansion of the triangle contribution stops at ${\cal O}(\pt^2)$.

Given the reducible structure of the double-triangle diagrams
(fig.\ref{fig:dia}$(g)$), an exact result for the double-triangle contribution
can be derived in terms of products of one-loop Passarino-Veltman functions
\cite{Passarino:1978jh}.     
Explicit expressions for this contribution are presented in
appendix \ref{app:due}. Although we write the amplitude using a different
tensorial structure with respect to ref.\cite{Davies:2020drs} we checked,
using the relations between the two tensorial structures reported in appendix
\ref{app:uno}, that our result is in agreement with the one presented
in ref.\cite{Hasselhuhn:2016rqt}.

The box contribution (fig.\ref{fig:dia}$(f)$) was computed evaluating
the two-loop multi-scale Feynman integrals via two different expansions:
a LME up to and including $\mathcal{O}(1/\mt^6)$ terms, and  an
expansion in the transverse momentum up to and including
${\cal O}(\pt^4)$ terms.
The former expansion was used as ``control'' expansion of the latter.
Indeed, the $\pt$-expanded result actually ``contains'' the LME one. The LME
differs from the expansion in $\pt$ by the fact that $\hat{s}$ is
treated as a small parameter with respect to $\mt^2$, and not on the same
footing as in the latter case. This implies that if the $\pt$-expanded result is
further expanded in terms of the  $\hat{s}/\mt^2$ ratio the LME result has to
be recovered. This way, we were able to reproduce, at the analytic level,
our LME result.

We conclude this section outlining some technical details concerning our
computation.  We generated the amplitudes using \texttt{FeynArts} \cite{Hahn:2000kx} and
contracted them with the projectors as defined in appendix \ref{app:uno}
using \texttt{FeynCalc }\cite{Mertig:1990an,Shtabovenko:2016sxi} and in-house
Mathematica routines.  We used  dimensional regularization and
the rule for the contraction of two epsilon tensors written in terms of
the determinant of $n$-dimensional metric tensors. This is not a consistent
procedure and  needs to be corrected. A correction term should be added
\cite{Larin:1993tq} to the form factors computed as described
above, $\amp^{(1,ndr)}_i$, namely
\beq
\amp^{(1)}_i = \amp^{(1,ndr)}_i -\frac{\as}{\pi} C_F \amp_i^{(0)}~.
\label{eq:larin}
\eeq
In order to check eq.(\ref{eq:larin}), following ref.\cite{Degrassi:2011vq}
we bypassed the problem of the treatment of 
$\gamma_5$ in dimensional regularization computing the amplitude via
a LME working in 4 dimension, employing the Background Field Method (BFM)
\cite{Abbott:1980hw} and using as regularization scheme  the Pauli-Villars
method. This result was compared with the LME evaluation of
$\amp^{(1,ndr)}_i$, finding that the difference between the two
evaluations was indeed given by the second term on the right-hand-side of
eq.(\ref{eq:larin}).

After the contraction of the epsilon tensors the diagrams were expanded as
described in section \ref{sec:tre}. They were reduced to MIs
using \texttt{FIRE} \cite{Smirnov:2014hma} and \texttt{LiteRed} \cite{Lee:2013mka}. The
resulting MIs were exactly the same as previously found for di-Higgs
production \cite{Bonciani:2018omm}. Nearly all of them are expressed
in terms of generalised harmonic polylogarithms with the exception of
two elliptic integrals \cite{vonManteuffel:2017hms, Bonciani:2018uvv}.
The top quark mass was renormalized in the onshell scheme\footnote{Different choices
  for the renormalized top mass can be easily implemented in our calculation.}
 and the IR poles were subtracted as in ref.\cite{Degrassi:2016vss}.

\section{NLO results} \label{sec:sei}
We now present our numerical results for the virtual corrections.
We have implemented our results into a \texttt{FORTRAN} programme. 
For the evaluation of the generalised harmonic polylogarithms we use 
the code \texttt{handyG} \cite{Naterop:2019xaf}, while 
the elliptic integrals are evaluated using the routines of
ref.\cite{Bonciani:2018uvv}.
In order to facilitate the comparison of our results with the ones
presented in the literature,  we define the finite part of the virtual corrections
as in
 ref.\cite{Davies:2020drs}\footnote{Our definition of the matrix elements
   differs by a factor of
   $\frac{1}{\hat{s}}$ from ref.\cite{Davies:2020drs}, \textit{cf}. also
   appendix \ref{app:uno}.}
\begin{equation}
\begin{split}
  \mathcal{V}_{fin}&=\frac{G_F^2 m_Z^2}{16}\left(\frac{\as}{\pi}\right)^2
  \left[ \sum_{i} \left|\mathcal{A}_i^{(0)} \right|^2\frac{C_A}{2}\left(\pi^2-
    \log^2\left(\frac{\mu_R^2}{\hat{s}}\right)\right)\right. \\
  & \left. +2\sum_i\text{Re}\left[\mathcal{A}_i^{(0)}\left(\mathcal{A}_i^{(1)}\right)^*\right]\right]\,
  \label{eq:vfin}
\end{split}
\end{equation}
and in the numerical evaluation of eq.(\ref{eq:vfin}) we fixed
$\mu_R= \sqrt{\hat{s}}$.

First, both the triangle and box LME contributions to $\mathcal{A}_i^{(1)}$ up
to $\mathcal{O}(1/\mt^6)$ terms  were checked, at the analytic level, against
the results of refs.\cite{Hasselhuhn:2016rqt,Davies:2020drs} finding perfect
agreement. Then,  the $\pt$-expanded results for low $M_{ZH}$ were
confronted numerically with the LME ones, finding a  good numerical agreement.
We recall that, at the same order in the expansion, 
the $\pt$-expanded terms are more accurate than
the LME ones, although computationally more demanding.

In ref.\cite{Chen:2020gae} a numerical evaluation of eq.(\ref{eq:vfin})
was presented. In that reference the exact NLO amplitude was reduced to a
set of MIs that were evaluated numerically using the code \texttt{pySecDec}
\cite{Borowka:2017idc,Borowka:2018goh}. Table 3 of that reference presents
the numerical results\footnote{The values in table 3 of ref.\cite{Chen:2020gae}
  are defined as $V_{fin} 4/(\as^2 \alpha^2)$.} for various points in the
phase space. For the four points in that table lying within the range of
validity of our expansion we find a difference with respect to our results of
 less than 1 permille in 3 cases and reaching the 1 permille level in one case, similarly to what we find at LO.
It should be noticed that  small differences
on the permille level can be explained not only by the different
approaches (exact vs. $\pt$-expanded) but also by the fact that
in ref.\cite{Chen:2020gae} the $\mz^2/\mt^2$ and $\mh^2/\mt^2$ ratios were
approximated by a ratio of two integer numbers.

\begin{figure}[th]
\includegraphics[width=\textwidth]{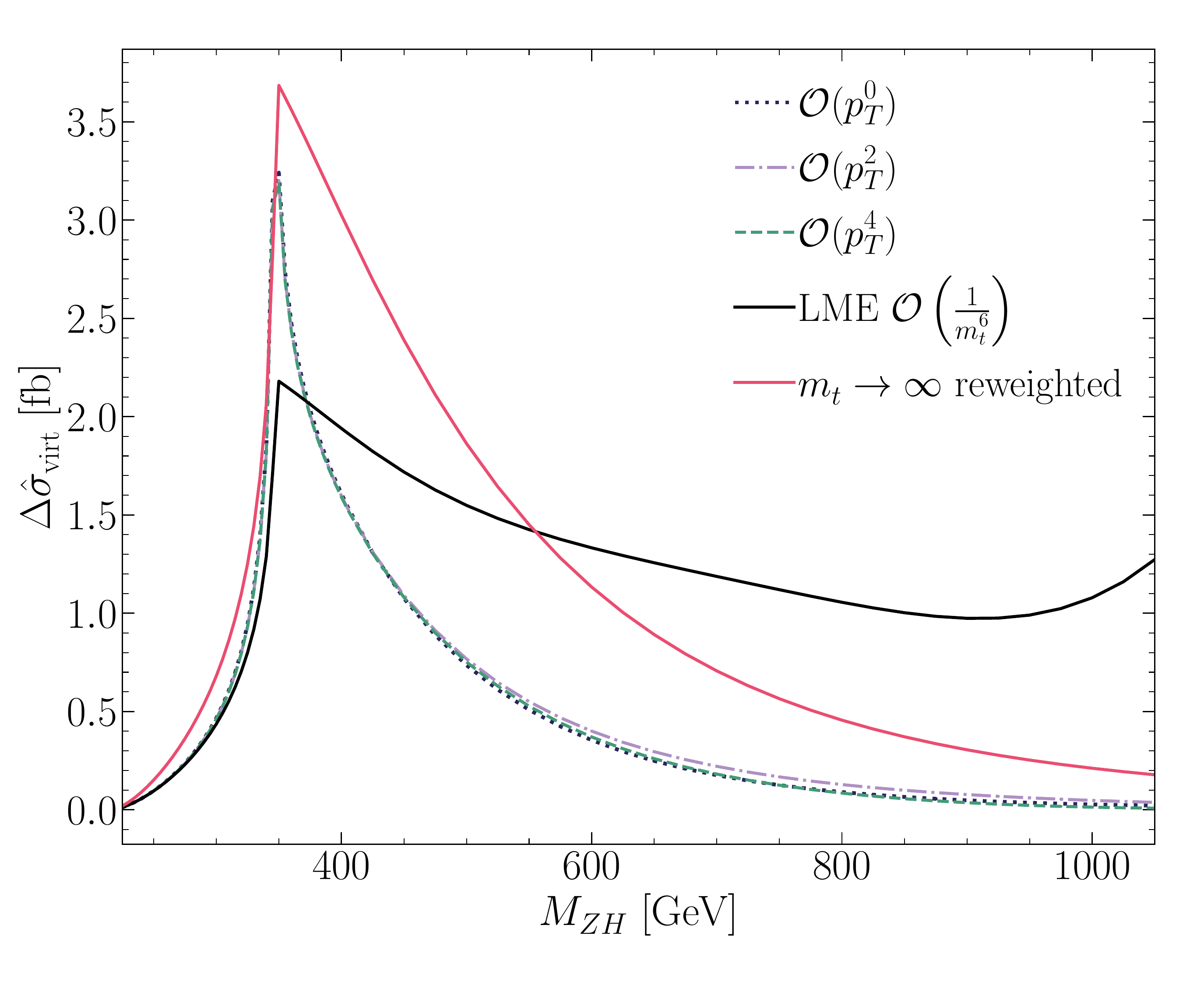}
\caption{$\Delta \hat{\sigma}_{virt}$ defined by eq.\eqref{eq:deltasigma}, shown as a function of $M_{ZH}$. The various orders of the $\pt$-expansion are plotted as dashed lines, while the black and red continuous lines stand for the LME and  reweighted $m_t \rightarrow \infty$ results, respectively.}
  \label{fig:deltasigma}
\end{figure}
In order to present our results we define a virtual part of the partonic cross section
from the finite part of the virtual corrections in eq.\eqref{eq:vfin} by
\begin{equation}
  \Delta \hat{\sigma}_{virt}=
  \int_{\hat{t}^-}^{\hat{t}^+} d\hat{t}
  \frac{\alpha_s}{16\pi^2}\frac{1}{\hat{s}^2}\mathcal{V}_{fin}\,
  \label{eq:deltasigma}
\end{equation}
and  show it in fig.\ref{fig:deltasigma}. The dashed lines in the
plot show the different orders in our expansion. 
For all parts of the matrix elements we use the best results
available, i.e.~both $\mathcal{A}^{(0)}$ and the double-triangle
contribution are evaluated  exactly, while for
$\mathcal{A}^{(1)}$ we use the various orders in the $\pt$-expansion.
For comparison, we show the results where
$\mathcal{A}^{(1)}$ is replaced by the one computed in LME up to
$\mathcal{O}(1/m_t^6)$ (full black line), which as mentioned before is valid
up to $M_{ZH}< 2 \mt$. We see that within the validity of the LME our
results agree well with it.
Furthermore, we show the results in the infinite top
mass limit reweighted by the full amplitudes squared (full red line), corresponding to the
approach of ref.\cite{Altenkamp:2012sx}, keeping though the double triangle
contribution in full top mass dependence. 
Differently from the LME line, the $\mt \to \infty$ reweighted one
shows a behaviour, for  $M_{ZH} \gtrsim 400\gev$, similar to the behaviour of
the $\pt$ lines. Still,   the difference
between the reweighted result and the $\pt$-expanded ones is  significant.
The $\pt$-expanded results show
very good convergence.  The zero order in our expansion agrees
extremely well with the higher orders in the expansion, and all the
three results are very close up to $M_{ZH} \sim 500\gev$.

Finally, we note that the evaluation of $\mathcal{V}_{fin}$ requires a
running time per phase space point less than
one second. In addition, the integration over the $\hat{t}$ variable
 in eq.(\ref{eq:deltasigma}) converges very well, such that 
 fig.\ref{fig:deltasigma} could be produced on a standard laptop in a
 few hours. Thus, our computation of the two-loop virtual corrections
 in $gg \to ZH$ is suitable to be used within a Monte Carlo code.

\section{Conclusion \label{sec:conclusion}}
In this paper, we computed  the two-loop NLO virtual corrections to the 
$gg \to ZH$ process. Among the two-loop Feynman diagrams contributing
to the process,  the ones belonging to the triangle and
double-triangle topology were computed exactly. The ones  belonging 
to the box topology, which contain multiscale integrals, were evaluated via an
expansion in the $Z$ transverse momentum. This novel approach of
computing a process in the forward kinematics
was originally proposed in ref.\cite{Bonciani:2018omm} for
double Higgs production where the particles in the final state have
the same mass. In this paper,  we  extended
the method to the more general case of two different masses in the
final state and to a process whose amplitude is not symmetric
under the  $\hat{t}\leftrightarrow \hat{u}$ exchange.

The result of the evaluation of the box contribution is expressed,
both at one- and two-loop level, in terms of the
same set of MIs that was found in ref.\cite{Bonciani:2018omm} for
double Higgs production.  The two-loop MIs  can be all
expressed in terms of generalised harmonic polylogarithms with the
exception of two elliptic integrals.

As we have shown explicitly at the LO, the range of validity of our
computation covers values of the invariant mass
$M_{ZH}\lesssim 750\text{ GeV}$ corresponding to 98.5\% of the phase
space at LHC energies.  We showed that few terms in our
expansion were sufficient to obtain an incredible good agreement with the
numerical evalution of ${\mathcal V}_{fin}$ presented in ref.\cite{Chen:2020gae},
at the level of a permille or less difference between our analytic result and the numerical one.

The advantage of our analytic approach compared to the numerical
calculation is also in the computing time. With an average evaluation
time of half a second per phase space point, an inclusion into a Monte Carlo
programme is realistic. Due to the flexibility of our analytic
results, an application to beyond-the-Standard Model is certainly
possible.

Finally, we remark that our calculation complements
nicely the results obtained in ref.\cite{Davies:2020drs} using a high-energy
expansion, that according to the authors provides precise results for
$\pt \gtrsim 200\gev$. The merging of the two analyses is going to provide
a result that covers the whole phase space, can be easily implemented into a
Monte Carlo code and  presents the flexibility of an analytic calculation.

\section*{Acknowledgements}
We are indebted to R.~Bonciani for his contribution during the first stage of
this work.
We thank G.~Heinrich and J.~Schlenk for help with the comparison with
their results, E.~Bagnaschi for help with optimising the numerical
evaluation of the GPLs, and M.~Kraus for discussions.  
The work of G.D. was partially supported by the Italian Ministry of Research
(MUR) under grant PRIN 20172LNEEZ. The work of
P.P.G. has received financial support from Xunta de Galicia (Centro
singular de investigaci\'on de Galicia accreditation 2019-2022), by
European Union ERDF, and by ``Mar\'ia de Maeztu" Units of Excellence
program MDM-2016-0692 and the Spanish Research State Agency; L.A 's research is funded by the Deutsche Forschungsgemeinschaft (DFG, German Research Foundation) - Projektnummer 417533893/GRK2575  ``Rethinking Quantum Field Theory".

\begin{appendletterA}
\section{Orthogonal Projectors in $gg \to ZH$}
\label{app:uno}
In this appendix we present
the explicit expressions of the projectors  $\mathcal{P}_i^{\mu\nu\rho}$
appearing in eq.(\ref{eq:amp}). The projectors are all normalized to 1.
They  are:
\bea
\mathcal{P}_1^{\mu\nu\rho}&=&\frac{\mz}{\sqrt{2}s'\pt^2}\biggl[ p_1^\nu
\epsilon^{\mu\rho p_1 p_2}-p_2^\mu \epsilon^{\nu\rho p_1 p_2}+
q_t^\mu \epsilon^{\nu\rho p_2 p_3}\\
&+&q_u^\nu \epsilon^{\mu\rho p_1 p_3}+s' \epsilon^{\mu\nu\rho p_2}-s'
\epsilon^{\mu\nu\rho p_1}\biggr],\\
\mathcal{P}_2^{\mu\nu\rho}&=&\frac{1}{\sqrt{2}s'\pt}
\biggl[ q_u^\nu \epsilon^{\mu\rho p_1 p_3}+q_t^\mu\epsilon^{\nu\rho p_2 p_3}\biggr],\\
\mathcal{P}_3^{\mu\nu\rho}&=&\frac{\sqrt{3}}{2s'\pt}
\biggl[ s' \epsilon^{\mu\nu\rho p_1}+s' \epsilon^{\mu\nu\rho p_2}-
p_1^\nu\epsilon^{\mu\rho p_1 p_2}-p_2^\mu \epsilon^{\nu\rho p_1 p_2} \nn \\
&+& \left( q_u^\nu \epsilon^{\mu\rho p_1 p_3} -q_t^\mu \epsilon^{\nu\rho p_2 p_3}
\right) \left(\frac13 + \frac{\mz^2}{\pt^2} \right) \nn\\ 
&+& \frac{\mz^2}{\pt^2}\left( q_t^\mu\epsilon^{\nu\rho p_2 p_1}-
q_u^\nu\epsilon^{\mu\rho p_1 p_2} \right) \biggr], \\
\mathcal{P}_4^{\mu\nu\rho}&=&\frac{\mz}{\sqrt{2}s'\pt^2}
\biggl[ q_t^\mu(\epsilon^{\nu\rho p_2 p_1}-\epsilon^{\nu\rho p_2 p_3})-
q_u^\nu (\epsilon^{\mu\rho p_1 p_2}-\epsilon^{\mu\rho p_1 p_3})\biggr], \\
\mathcal{P}_5^{\mu\nu\rho}&=&\frac{1}{\sqrt{6}s'\pt}
\biggl[ q_t^\mu\epsilon^{\nu\rho p_2 p_3}-q_u^\nu \epsilon^{\mu\rho p_1 p_3}\biggr],\\
\mathcal{P}_6^{\mu\nu\rho}&=&\frac1{s'\pt}\biggl[
g^{\mu\nu} \epsilon^{\rho p_1 p_2 p_3}+s' \epsilon^{\mu\nu\rho p_3}+
p_1^\nu \epsilon^{\mu\rho p_2 p_3}-p_2^\mu \epsilon^{\nu\rho p_1 p_3}-
\frac{s'}2 \epsilon^{\mu\nu\rho p_2}  \nn \\
&+&\frac12 \left( p_1^\nu\epsilon^{\mu\rho p_1 p_2}+p_2^\mu \epsilon^{\nu\rho p_1 p_2}+
q_u^\nu \epsilon^{\mu\rho p_1 p_3}-q_t^\mu \epsilon^{\nu\rho p_2 p_3}
- s' \epsilon^{\mu\nu\rho p_1} \right) \nn \\
&+& \frac{\mz^2}{2\pt^2} (q_t^\mu\epsilon^{\nu\rho p_2 p_1}-
q_u^\nu\epsilon^{\mu\rho p_1 p_2}+ q_u^\nu \epsilon^{\mu\rho p_1 p_3}-
q_t^\mu \epsilon^{\nu\rho p_2 p_3})\biggr]~,
\eea
where we defined $q_t^\mu=(p_3^\mu-\frac{t'}{s'} p_2^\mu)$ and $q_u^\nu=(p_3^\nu-\frac{u'}{s'} p_1^\nu)$ and we used the shorthand notation $\epsilon^{\mu\nu\rho p_2}\equiv \epsilon^{\mu\nu\rho\sigma}p_2^\sigma$.

Using these projectors we obtained the relations between the form factors
$\amp_i$ defined in in eq.(\ref{eq:amp}) and those defined in
section 2 of ref.\cite{Davies:2020drs}:
\bea
\amp_1&=&\frac{\pt^2}{2\sqrt{2}\mz(\pt^2+\mz^2)}
\biggl[ (t'+u')F_{12}^+-(t'-u')F_{12}^- \biggr], \\
\amp_2&=&-\frac{\pt}{2\sqrt{2}(\pt^2+\mz^2)} \biggl[
(t'+u')F_{12}^+-(t'-u')F_{12}^- \nn \\
&-&\frac{\pt^2+\mz^2}{2 s'}((t'+u')F_{3}^+-(t'-u')F_{3}^-) \biggr],\\
\amp_3&=&\frac{\pt}{2\sqrt{3}(\pt^2+\mz^2)} \biggl[
(t'+u')F_{12}^--(t'-u')F_{12}^+ \nn \\
&+& (\pt^2+\mz^2)(F_{2}^-+F_4) \biggr],\\
\amp_4&=&-\frac{\mz}{2\sqrt{2}(\pt^2+\mz^2)} \biggr[
(t'+u')F_{12}^--(t'-u')F_{12}^+ \nn \\
&+&(\pt^2+\mz^2)\left( (1-\frac{\pt^2}{\mz^2})F_2^-+2F_4 \right) \biggl],\\
\amp_5&=&\frac{\pt}{2\sqrt{6}(\pt^2+\mz^2)} \biggr[
(t'+u')F_{12}^--(t'-u')F_{12}^+ \nn \\
  &+& (\pt^2+\mz^2) \left(
  4(F_2^-+F_4)+\frac{3}{2s'}\left( (t'+u')F_{3}^--(t'-u')F_{3}^+ \right) \right)
  \biggr],\nn\\
&& \\
\amp_6&=&\frac{\pt}{2}F_4.
\eea

\end{appendletterA}

\begin{appendletterB}
\section{Two-loop Results}
\label{app:due}

The NLO amplitude can be written in terms of three contributions,
namely the two-loop 1PI triangle, the two-loop 1PI box and the
reducible double-triangle diagrams,
\begin{equation}
  \mathcal{A}_{i}^{(1)} = \mathcal{A}_{i}^{(1, \triangle) } +
  \mathcal{A}_{i}^{(1, \square) } + \mathcal{A}_{i}^{(1, \bowtie) }~.
\end{equation}
We present here  the exact results for the  
double-triangle and triangle contributions to all the form factors. We find
\bea
\mathcal{A}_{1}^{(1, \bowtie)} &=& -\frac{\mt^2 \pt^2}{4 \sqrt{2}~ \mz \left(\mz^2+\pt^2\right)^2} \Biggl[ F_t(\hat{t}) \left(G_t(\hat{t},\hat{u})-G_b(\hat{t},\hat{u})\right)+ (\hat{t} \leftrightarrow \hat{u}) \Biggr], \\
\mathcal{A}_{2}^{(1, \bowtie)} &=& \frac{\mt^2 \pt}{4 \sqrt{2} \left(\mz^2+\pt^2\right)^2}  \Biggl[ F_t(\hat{t}) \left(G_t(\hat{t},\hat{u})-G_b(\hat{t},\hat{u})\right)+ (\hat{t} \leftrightarrow \hat{u})\Biggr], \\
\mathcal{A}_{3}^{(1, \bowtie)} &=& \frac{\mt^2 \pt}{4 \sqrt{3} ~ \hat{s} \left(\mz^2+\pt^2\right)^2} \Biggl[\left(\mh^2-\hat{t}\right) F_t(\hat{t}) \left(G_t(\hat{t},\hat{u})-G_b(\hat{t},\hat{u})\right)- (\hat{t} \leftrightarrow \hat{u})\Biggr], \nn\\
&&\\
\mathcal{A}_{4}^{(1, \bowtie)} &=& -\frac{\mt^2}{4 \sqrt{2} ~ \mz  \hat{s}^2 \left(\mz^2+\pt^2\right)^2} \Biggl[ \Bigl(\mz^2 \left(\mh^2-\hat{t}\right)^2\nn \\
& -&\hat{t} \left(\mz^2-\hat{u}\right)^2\Bigr) F_t(\hat{t})
   \left(G_t(\hat{t},\hat{u})-G_b(\hat{t},\hat{u})\right) 
 - (\hat{t} \leftrightarrow \hat{u})   \Biggr], \\
   \mathcal{A}_{5}^{(1, \bowtie)} &=& -\frac{\mt^2 \pt}{4 \sqrt{6} ~ \hat{s} \left(\mz^2+\pt^2\right)^2} \Biggl[  \left(4 \mz^2-\hat{s}-4
   \hat{u}\right) F_t(\hat{t}) \left(G_t(\hat{t},\hat{u})-G_b(\hat{t},\hat{u})\right)\nn \\
  &-& (\hat{t} \leftrightarrow \hat{u}) \Biggr], \\
   \mathcal{A}_{6}^{(1, \bowtie)} &=&  0,
\eea
where
\bea
  F_t(\hat{t}) &=& \frac{1}{\left(\mh^2-\hat{t}\right)^2}
  \Biggl[2 \hat{t} \Bigl( B_0\left(\hat{t},\mt^2,\mt^2\right)-
    B_0\left(\mh^2,\mt^2,\mt^2\right) \Bigr)\nn \\
& +&\left(\mh^2-\hat{t}\right)
    \Bigl(\left(\mh^2-4 \mt^2-\hat{t}\right)
    C_0\left(0,\mh^2,\hat{t},\mt^2,\mt^2,\mt^2\right)-2\Bigr)\Biggr], \nn\\
    &&\\
G_x(\hat{t},\hat{u})   &=& \left(\mz^2-\hat{u}\right)  \Biggl[\mz^2
  \Bigl( B_0\left(\hat{t},m_x^2,m_x^2\right) -
  B_0\left(\mz^2,m_x^2,m_x^2\right) \Bigr) \nn\\
  & +&\left(\hat{t}-\mz^2\right) \Bigl(2 m_x^2
   C_0\left(0,\hat{t},\mz^2,m_x^2,m_x^2,m_x^2\right)+1\Bigr)\Biggr].
\eea

Instead, for the triangle diagrams, we obtain
\bea
  \mathcal{A}_{1}^{(1, \triangle) } &=& \frac{ \pt^2 ~
    ( \hat{s}-\dm)}{4 \sqrt{2} \mz } \frac{\mathcal{K}_t^{(2l)}}{
    \left(\pt^2+\mz^2\right)}, \\
  \mathcal{A}_{2}^{(1, \triangle) } &=& - \frac{ \pt ~ ( \hat{s}-\dm)}{4 \sqrt{2} }
  \frac{\mathcal{K}_t^{(2l)}}{ \left(\pt^2+\mz^2\right)},\\
  \mathcal{A}_{3}^{(1, \triangle) } &=&  \frac{ \pt ~ (\hat{t}-\hat{u})}{4 \sqrt{3} }
  \frac{\mathcal{K}_t^{(2l)}}{\left(\pt^2+\mz^2\right)}, \\
  \mathcal{A}_{4}^{(1, \triangle) } &=& -\frac{ \mz ~ (\hat{t}-\hat{u})}{4 \sqrt{2} }
  \frac{\mathcal{K}_t^{(2l)}}{ \left(\pt^2+\mz^2\right)}, \\
  \mathcal{A}_{5}^{(1, \triangle) } &=& -\frac{ \pt ~ (\hat{t}-\hat{u})}{4 \sqrt{6} }
  \frac{\mathcal{K}_t^{(2l)}}{\left(\pt^2+\mz^2\right)}, \\
\mathcal{A}_{6}^{(1, \triangle) } &=& 0,
\eea
where the $\mathcal{K}_t^{(2l)}$ function is defined in eq.(4.11) of
ref.\cite{Aglietti:2006tp}.

We do not show the explicit results for the $\pt$-expansion of the
two-loop box diagrams, since the analytic expressions are very
lengthy, even for the lowest order term of the expansion.

\end{appendletterB}

\bibliographystyle{utphys}
\bibliography{ADGGV}

\end{document}